\documentclass{bioinfo}
\copyrightyear{2005}
\pubyear{2005}

\begin{document}
\firstpage{1}

\title[bwa-meth]{Fast and accurate alignment of long bisulfite-seq reads}
\author[Pedersen \textit{et~al}]{
Brent S. Pedersen\,$^{1,}
\footnote{To whom correspondence should be addressed.  Email: bpederse@gmail.com}$,
Kenneth Eyring\,$^{1}$,
Subhajyoti De\,$^{1,2}$,
Ivana V. Yang\,$^{1}$
and David A. Schwartz\,$^1$%
}
\address{
    $^{1}$Department of Medicine, University of Colorado Denver, School of Medicine, Denver, Colorado, USA. 80045 \\
    $^{2}$University of Colorado Cancer Center, Molecular Oncology Program,
    Aurora, Colorado, USA
}

\history{Received on XXXXX; revised on XXXXX; accepted on XXXXX}

\editor{Associate Editor: XXXXXXX}

\maketitle

\begin{abstract}


\section{Summary:}
We introduce a new tool, \textit{bwa-meth}, to align bisulfite-treated
sequences and compare it to existing aligners. We show that it is fast
and accurate even without quality-trimming and 
the output is immediately usable by downstream tools.
We gauge accuracy by comparing the number of on and off-target
reads from a targeted sequencing project and by simulations.

\section{Availability and Implementation:}
The benchmarking scripts and the \textit{bwa-meth} software are available at
https://github/com/brentp/bwa-meth/ under the MIT License.

\section{Contact:} \href{bpederse@gmail.com}{bpederse@gmail.com}
\section{Supplementary information:} 
Supplemental Information I
\end{abstract}

\section{Introduction}
Bisulfite sequencing (BS-Seq) is a common way to explore DNA methylation.
As a result, software 
has been developed to map sequence reads treated with bisulfite to a reference genome \citep{frithlast,methylcoder,gsnap,krueger2011,bsmap,bsmooth}.
Previous studies have compared alignment statistics on 
real \citep{methylcoder,bsmap,shrestha} and simulated \citep{frithlast} reads,
however most of these are limited by knowledge of the ground-truth and assumptions of
the simulation, respectively.

Here, we present an analysis of current BS-Seq mappers including the "four-base" aligners
Last \citep{frithlast}, GSNAP \citep{gsnap} and BSMAP \citep{bsmap} as well as the
"three-base" aligners BSmooth \citep{bsmooth}, Bison, and Bismark \citep{krueger2011} which
perform \emph{in silico} conversion of cytosines to thymines. In addtion, we introduce our
own, simple three-base aligner that wraps \textit{BWA mem} \citep{bwamem}.

The comparison is performed on 100-base paired-end reads
which are of modest length by current standards, but, to our knowledge, longer than
utilized in any comparison. We hypothesized that long, paired reads, with up
to 200 bases from the same genomic region, could change the decision on which
alignment method performed the best and that aligners which relied on global
alignment might have reduced performance on longer reads.
We found limitations to exisiting aligners including the writing of large temporary
files, required quality-trimming, high memory-use, long run-time, output that was not
suitable for consumption by traditional tools, or some combination of these
inconveniences. We wrote
a BS-Seq aligner based on BWA mem \citep{bwamem} to address these
limitations. This new aligner, 
\textit{bwa-meth}, allows indels, local alignments, and it never writes a
temporary-file of the reads to disk, instead streaming the \emph{in silico} converted
reads directly to the aligner and streaming the alignments directly to a well-formatted
alignment file suitable for use in downstream tools. In addition, it works well without
quality-trimming, thereby reducing storage requirements 3-fold by not creating
quality-trimmed or converted reads.

\section{Approach}
In order to determine the accuracy of an aligner,
we utilize a dataset from Agilent's SureSelect Mouse Methyl-Seq kit which
captures about
99 million bases from CpG dense regions in the mouse genome (a similar
approach is available for human regions).
We evaulate an aligner by the number of reads in the capture area as compared
to outside of the capture area. While there will be off-target capture, all
aligners are subject to the same assumptions. With those constraints, we can
plot a receiver operating curve (ROC) with true positives as reads within
and false positives as reads outside of the target regions.
\citealp{shrestha} performed an unbiased analysis on 35 base reads sequenced from
the X chromosome and called a read as accurately mapped if it aligned
to the X chromosome; here, we map to the entire genome with 100-base paired-end reads.

In addition, we map 100-base paired-end indel-free reads simulated using
Sherman (v0.1.6), the software from the authors of Bismark. All data were
aligned to mouse genome version \textit{mm10}.

\begin{methods}
\section{Methods}
We aligned real and simulated data, both as-is and trimmed by quality (using Trim Galore
[\href{http://www.bioinformatics.babraham.ac.uk/projects/trim\_galore/}{http://www.bioinformatics.babraham.ac.uk/projects/trim\_galore/}]
default parameters), using the software and versions in Supplemental Table 1.
We evaluated a few parameters for each method and report only the
best-performing here. Likely, every aligner could show improved results with
exhaustive search of the parameters, but this is a representation of
reasonable selections.
We considered a real read to be on-target if it was within 1001 bases
of a target region.

We designed \textit{bwa-meth} for paired-end reads from the directional
protocol but it can align single-end reads. \textit{bwa-meth} outputs alignments
directly to a BAM file that is usable even by those tools that require
coordinate-sorted alignments and read-groups. Since it consists
of fewer than 600 lines of code and runs
quickly, it can be used as a platform to test other optimizations. For example
in Supplemental Figure 7, we show a feature that further reduces the
number of off-target reads by considering only the single strand targeted by
the SureSelect protocol.

Although our comparisons are on an inbred mouse strain with few polymorphisms
we expect the results will hold even with insertions and deletions due to
our use of BWA mem \citep{bwamem}.

\end{methods}

\begin{figure}[!tpb]
    \centerline{\includegraphics[width=86mm]{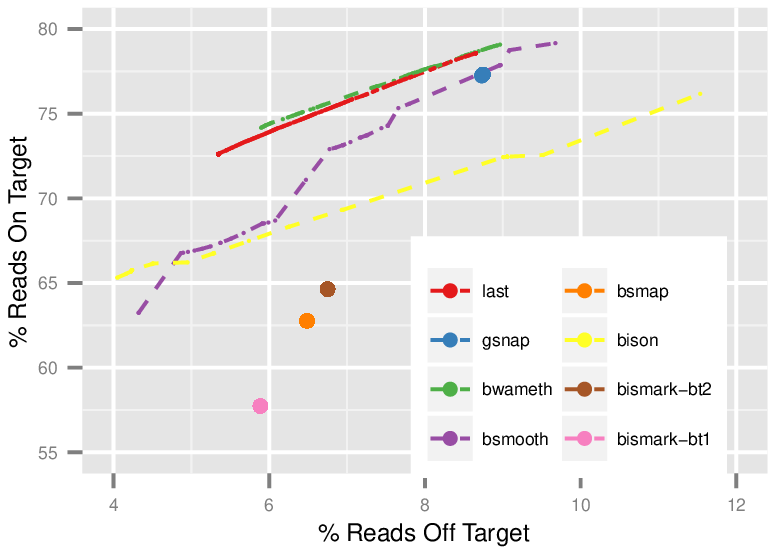}}
    \caption{Percent of un-trimmed paired-end, 100-base reads on (y) and off (x)
    target for the tested aligners. Aligners that report mapping quality are
shown as connected dots for each quality cut-off. Reads are limited to those
considered as primary, mapped alignments by the aligner. Note that, as shown
in the supplement, many aligners have better performance on trimmed reads.
}\label{fig:01}
\end{figure}

\section{Discussion}

\subsection{Accuracy}
For the aligners that report a range of mapping quality
scores--an indicator of the aligner's confidence in the
alignment--we vary the score from 1 to the maximum, 255, to draw an ROC-like
curve showing the trade-off between sensitivity and specificity. For the
other aligners, we plot their single location. Figure \ref{fig:01} shows
the on and off-target reads for our real paired-end data. Last and 
\textit{bwa-meth} align the most reads on target with a low percent
of off-target reads, but Last can provide better control over the number
of off-target reads. On trimmed data, \textit{bwa-meth}, Last, Bison and
Bsmooth all perform similarly with Bison allowing very good control
over the number off-target reads (Supplement Figure 2).
Only Last, GSNAP, and \textit{bwa-meth} are relatively unaffected by quality-trimming.

Bismark performance varies depending on whether it uses bowtie version
1 or 2 for the backend but in either case, (as with all of the bowtie-backed
aligners) it performs better with trimmed reads.

A similar comparison is shown in Supplemental Figure 2 for trimmed data.
We also show the ROC figures for simulated data with errors in Supplemental
Figures 3 (original) and 4 (trimmed) and without errors in Supplemental 
Figures 5 and 6.
\textit{Bwa-meth} does out-perform all aligners for simulated data.

\subsection{Computational Resources}
We are more interested in the accuracy of a
method than the speed. In general the aligners are comparable in terms of
speed. While Bismark with bowtie1 is quite fast on the simulated data it
can not use multiple processes and so takes the longest user-time. Bsmap
takes the longest CPU time on the real data. We report the exact 
timings and maximum memory use in Supplemental Information.
Bsmap uses the least disk, never writing an index of the reference genome
and only writing the alignment files. All other aligners write an index of
the reference genome. Last, bsmooth, and bismark write additional copies of the
reads to disk.
\textit{Bwa-meth} avoids writing the \emph{in silico}
converted reads to disk by streaming them directly to the aligner and shows
nearly identical accuracy without read-trimming. The trimming and conversion
steps can increase data storage needs enough to be a burden in our experience. 

None of the programs used an inordinate amount of memory, however, due to
the parallelization strategy, Last did require about 10GB of shared memory
per process.

\section{Conclusion}
We have used simulated reads along with reads from a capture method that
targets CpG-rich regions to compare aligners.
We show that \textit{bwa-meth} is very accurate, even without quality trimming.
Aligners that use global alignment benefit more from quality trimming and may
also require an additional copy of the \emph{in silico} converted reads.
\textit{bwa-meth} is also fast and it outputs alignments to a sorted bam
that is immediately usable by downstream tools.

\section*{Acknowledgement}
We thank Devon Ryan for several helpful conversations.
\paragraph{Funding\textcolon} This was funded by R01 HL097163, R01 HL101251, 1I01BX001534, RC2 HL101715, N01 AI90052, and S10 RR031832.
%
%

\bibliographystyle{natbib}
    \bibliography{document}

\begin{thebibliography}{}

\bibitem[Frith {\em et~al.}(2012)Frith, Mori, and Asai]{frithlast}
Frith, M.~C., Mori, R., and Asai, K. (2012).
\newblock A mostly traditional approach improves alignment of
  bisulfite-converted {DNA}.
\newblock {\em Nucleic acids research\/}, {\bf 40}(13), e100--e100.

\bibitem[Hansen {\em et~al.}(2012)Hansen, Langmead, Irizarry, {\em
  et~al.}]{bsmooth}
Hansen, K.~D., Langmead, B., Irizarry, R.~A., {\em et~al.} (2012).
\newblock Bsmooth: from whole genome bisulfite sequencing reads to
  differentially methylated regions.
\newblock {\em Genome Biol\/}, {\bf 13}(10), R83.

\bibitem[Krueger and Andrews(2011)Krueger and Andrews]{krueger2011}
Krueger, F. and Andrews, S.~R. (2011).
\newblock Bismark: a flexible aligner and methylation caller for bisulfite-seq
  applications.
\newblock {\em Bioinformatics\/}, {\bf 27}(11), 1571--1572.

\bibitem[Li(2013)Li]{bwamem}
Li, H. (2013).
\newblock Aligning sequence reads, clone sequences and assembly contigs with
  bwa-mem.
\newblock {\em {arXiv} preprint arXiv:1303.3997\/}.

\bibitem[Pedersen {\em et~al.}(2011)Pedersen, Hsieh, Ibarra, and
  Fischer]{methylcoder}
Pedersen, B.~S., Hsieh, T.-F., Ibarra, C., and Fischer, R.~L. (2011).
\newblock {MethylCoder:} software pipeline for bisulfite-treated sequences.
\newblock {\em Bioinformatics\/}, {\bf 27}(17), 2435--2436.

\bibitem[Shrestha and Frith(2013)Shrestha and Frith]{shrestha}
Shrestha, A. M.~S. and Frith, M.~C. (2013).
\newblock An approximate bayesian approach for mapping paired-end dna reads to
  a reference genome.
\newblock {\em Bioinformatics\/}, {\bf 29}(8), 965--972.

\bibitem[Wu and Nacu(2010)Wu and Nacu]{gsnap}
Wu, T.~D. and Nacu, S. (2010).
\newblock Fast and {SNP-tolerant} detection of complex variants and splicing in
  short reads.
\newblock {\em Bioinformatics\/}, {\bf 26}(7), 873--881.

\bibitem[Xi and Li(2009)Xi and Li]{bsmap}
Xi, Y. and Li, W. (2009).
\newblock {BSMAP:} whole genome bisulfite sequence {MAPping} program.
\newblock {\em {BMC} bioinformatics\/}, {\bf 10}(1), 232.

\end{thebibliography}
\end{document}


\maketitle

\section{Advantages}
\textit{bwa-meth} provides a number of advantages:

1. ease of use -- a single script indexes the fasta, aligns the reads, and
   tabulates methylation. Default parameters work well.

2. speed -- due to the efficient parallelization by bwa, bwa-meth can run on
   as many CPU's are available on a single machine.

3. disk-usage -- bwa-meth can take either compressed or uncompressed files
   and the reads are streamed directly to the aligner, never written to disk. An
   aligner that requires trimmed reads and that writes the converted reads to
   disk will require 3X as much storage for just the raw sequence data.

4. useful output -- the BAM that results from bwa-meth is sorted, indexed and
   contains the proper mapping quality scores, alignment flags, and read-groups.
   In addition, the header @SQ lines are sorted so that tools like Picard and
   GATK will accept them as is.

5. simplicity -- we have carefully designed bwa-meth to be quite simple and
   the code readable so that enhancements are simple. For example, see \#6.

6. strand-specificity -- Devon Ryan,
   the author of Bison, noted that many capture methods target only the reverse
   strand -- commonly called the original bottom. Though this would be simple to
   add to any aligner, ours is the only one that currently supports this. See
   from the supplemental figure 5 for the reduction in off-target reads that results
   from only considering reads that align to the original bottom strand.

7. specificity and sensitivity -- while the difference between aligners in
   this regard is small, we have shown that bwa-meth consistently performs as
   the best or among the best. We believe this to be the case in general and
   not limited to the specific cases that we have shown.

8. accuracy without trimming -- the accuracy is due in part to bwa's local
   alignment. This is something that is not supported in bismark and means that
   we can achieve very good accuracy without performing additional trimming
   which would result in additional disk-usage and processing time.

\section{Aligners Compared}

\begin{table}[!htp]
\caption{Methods Compared}
{\begin{tabular}{lll}\\\toprule
software & version & arguments\\\midrule
bismark (bowtie1) & 0.11.1 & --gzip --bam -n 2 -l 24\\
bismark (bowtie2) & 0.11.1 & --gzip --bam -N 1 --score\_min L,-0.4,-0.5\\ 
bsmap & 2.74 & bsmap -v3 -m0 -x1000 -S42 -n0 -s12 -I1\\
bsmooth & fb3f7ef & --very-sensitive\\
bison & 0.3.0 & --directional --very-sensitive-local -N 1\\
bwa-meth & 0.10 & bwa-meth\\
gsnap & 2014-02-28 & --npaths 1 --quiet-if-excessive -k 15\\
last & 392 & last-bisulfite-paired.sh\\\bottomrule
\end{tabular}}
\end{table}
We used a modified version of the calling script for last available in the
github repository.

\section{Comparison On Real Data}

We compared 7 aligners on real data as described in the text. Supplemental
Figure 1 below shows the same data as Figure 1 from the main text.

\begin{figure}[H]
    \centerline{\includegraphics[width=125mm]{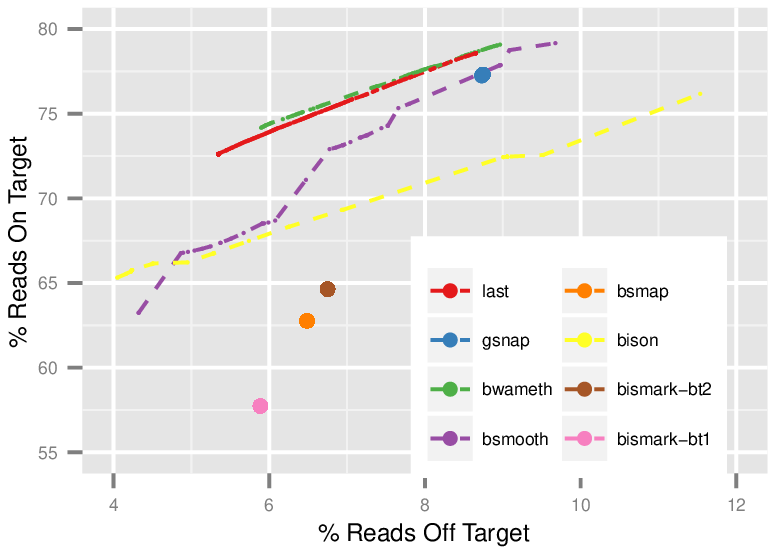}}
    \caption{Percent of paired-end, 100-base reads on (y) and off (x) target for
    the tested aligners. Aligners that report mapping quality are shown as lines
that span each quality cut-off. Reads are limited to those considered as primary, mapped alignments by the aligner. This is a color version of figure 1 from the paper}\label{suppfig:01}
\end{figure}

When we trim the reads with \emph{trim\_galore} before aligning, the result is shown below in Supplemental Figure 2. We report the percent of reads aligned relative
to the original count in the untrimmed data because we are interested in the
overall mapping rate.

\begin{figure}[H]
    \centerline{\includegraphics[width=125mm]{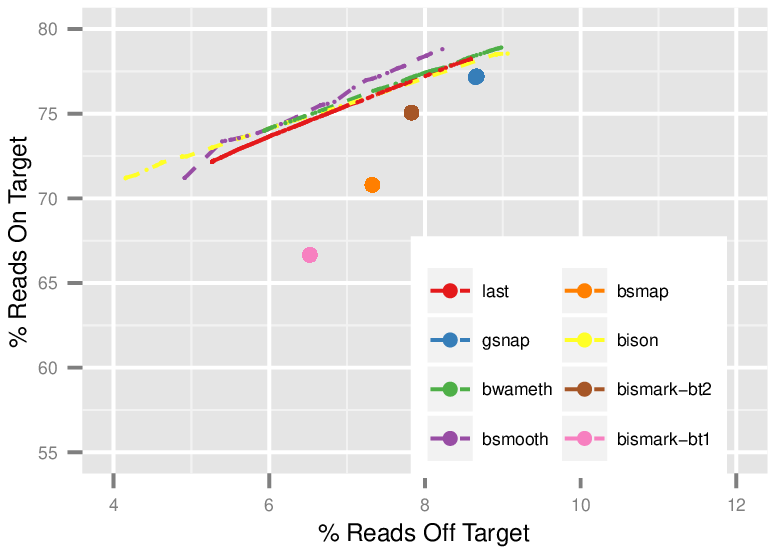}}
    \caption{Percent of paired-end, \emph{trimmed} 100-base reads on (y) and off (x) target for the tested aligners. Aligners that report mapping quality are shown as connected dots for each quality cut-off. Reads are limited to those considered as primary, mapped alignments by the aligner.}\label{suppfig:02}
\end{figure}

\subsection{Resources}

\begin{table}[H]
    \centering
    \caption{Resources on real data}
    \begin{tabular}{lllll} \hline
    trimmed & program & CPU time(min) & mem(GB) & dataset \\ \hline

no &    bis1 & 588.68 & 8.30 & real \\
no &    bis2 & 2067.23 & 9.88 & real \\
no &    bison & 1639.25 & 10.34 & real \\
no &    bsmap & 8498.52 & 23.06 & real \\
no &    bsmooth & 5469.96 & 9.94 & real \\
no &    bwa & 490.37 & 17.70 & real \\
no &    gsnap & 5309.48 & 12.73 & real \\
no &    last & 611.22 & 34.53 & real \\

yes &    bis1 & 581.60 & 8.28 & real \\
yes &    bis2 & 1921.00 & 9.64 & real \\
yes &    bison & 1433.06 & 10.06 & real \\
yes &    bsmap & 7259.29 & 23.19 & real \\
yes &    bsmooth & 5189.16 & 9.46 & real \\
yes &    bwa & 413.69 & 17.65 & real \\
yes &    gsnap & 4565.57 & 13.93 & real \\
yes &    last & 560.53 & 34.53 & real \\

    \end{tabular}
\end{table}

\section{Comparison On Simulated Data}
Paired-end, 100-base reads were simulated using the tool \emph{Sherman}
(http://www.bioinformatics.babraham.ac.uk/projects/sherman/) with 95\% of the
reads simulated from mm10 and 5\% of the reads from \emph{E. coli}. 
Reads were aligned with the parameters in Supplemental Table 1.
Reads from \emph{E. coli} aligning to \emph{mm10} were considered as
off-target reads. The Y-axis in the plot is the percent of the \emph{mm10}
reads. Exact parameters sent to Sherman are here:
https://github.com/brentp/bwa-meth/blob/master/compare/src/gen-simulated.sh

Note that these reads have no insertions or deletions. We simulated the 
\emph{E. coli} reads to be 98\% unmethylated and we used a per-base error
rate of 1\% for the simulations.

Supplemental Figure 3 below shows the result for these simulations.

\begin{figure}[H]
    \centerline{\includegraphics[width=125mm]{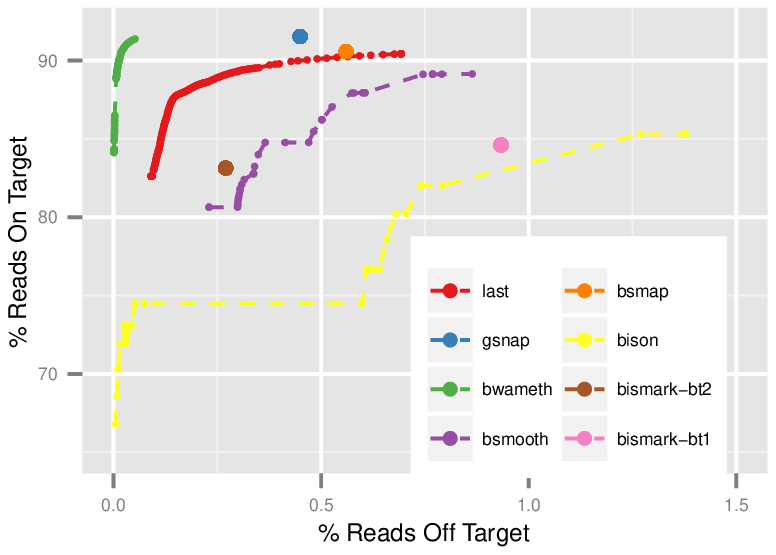}}
    \caption{Percent of paired-end, \emph{simulated} 100-base reads on (y) and off (x)
    target for the tested aligners. Aligners that report mapping quality are shown as
    connected dots for each quality cut-off. Reads are limited to those considered as
    primary, mapped alignments by the aligner.
}\label{suppfig:03}
\end{figure}

Supplemental Figure 4 below shows the result for trimmed and simulated reads.
Trimming removes some read-pairs if one or both reads had low-quality. We
report the percent of reads aligned relative to the original, 
un-trimmed number from \emph{mm10} since we are interested in the overall
mapping rate.

\begin{figure}[H]
    \centerline{\includegraphics[width=125mm]{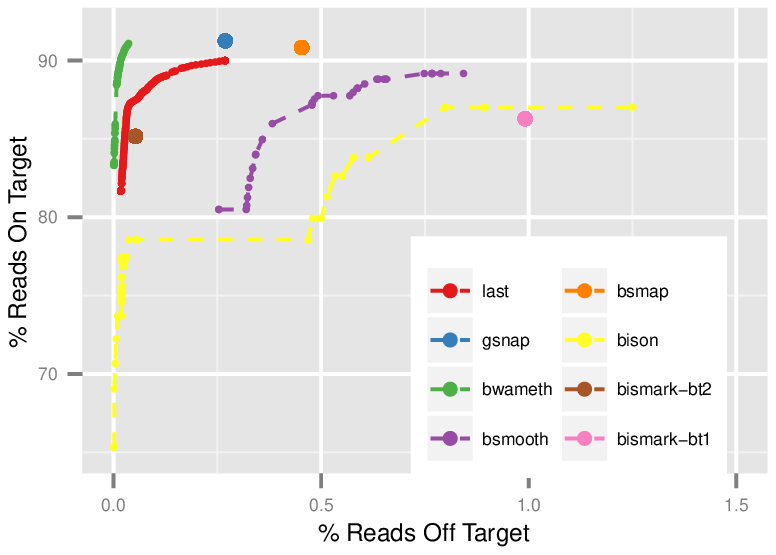}}
    \caption{Percent of paired-end, \emph{trimmed}, \emph{simulated} 100-base reads on (y) and off (x) target for the tested aligners. Aligners that report mapping quality are shown as connected dots for each quality cut-off. Reads are limited to those considered as primary, mapped alignments by the aligner.
}\label{suppfig:04}
\end{figure}

\subsection{Resources}

\begin{table}[H]
    \centering
    \caption{Resources on simulated data with 1\% error-rate
(programs run with more processes will use more memory)}
    \begin{tabular}{lllll} \hline
    trimmed & program & CPU time(min) & mem(GB) & dataset \\ \hline

no &    bis1 & 69.07 & 8.28 & sim \\
no &    bis2 & 271.13 & 9.66 & sim \\
no &    bison & 222.23 & 9.51 & sim \\
no &    bsmap & 227.35 & 22.99 & sim \\
no &    bsmooth & 501.97 & 9.52 & sim \\
no &    bwa & 145.56 & 19.40 & sim \\
no &    gsnap & 265.41 & 10.84 & sim \\
no &    last & 138.50 & 25.90 & sim \\

yes &    bis1 & 62.92 & 8.27 & sim \\
yes &    bis2 & 227.30 & 9.47 & sim \\
yes &    bison & 154.81 & 9.16 & sim \\
yes &    bsmap & 93.70 & 22.84 & sim \\
yes &    bsmooth & 461.71 & 9.31 & sim \\
yes &    bwa & 199.15 & 22.25 & sim \\
yes &    gsnap & 91.98 & 10.78 & sim \\
yes &    last & 113.80 & 25.90 & sim \\

    \end{tabular}
\end{table}

\section{Comparison On Error-Free Simulated Data}

Paired-end, error-free 100-base reads were simulated as above but
without contamination or errors.
Exact parameters sent to Sherman are here:
https://github.com/brentp/bwa-meth/blob/master/compare/src/gen-simulated.sh

Supplemental Figure 5 below shows the result for these simulations.
    
\begin{figure}[H]
    \centerline{\includegraphics[width=125mm]{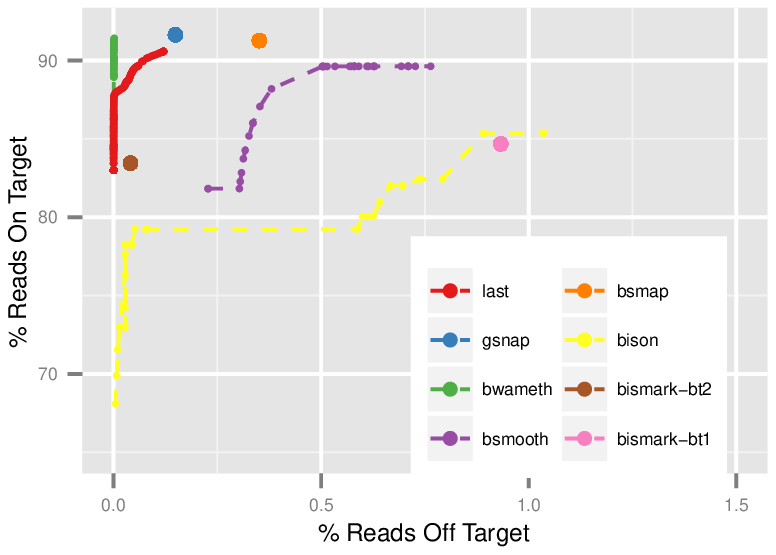}}
    \caption{Percent of paired-end, \emph{error-free, simulated} 100-base reads on (y) and off (x)
    target for the tested aligners. Aligners that report mapping quality are shown as
    connected dots for each quality cut-off. Reads are limited to those considered as
    primary, mapped alignments by the aligner.
}\label{suppfig:05}
\end{figure}

Supplemental Figure 6 below shows the result for trimmed and simulated reads.
Trimming removes some read-pairs if one or both reads had low-quality. We
report the percent of reads aligned relative to the original, 
un-trimmed number from \emph{mm10} since we are interested in the overall
mapping rate. 
 
\begin{figure}[H]
    \centerline{\includegraphics[width=125mm]{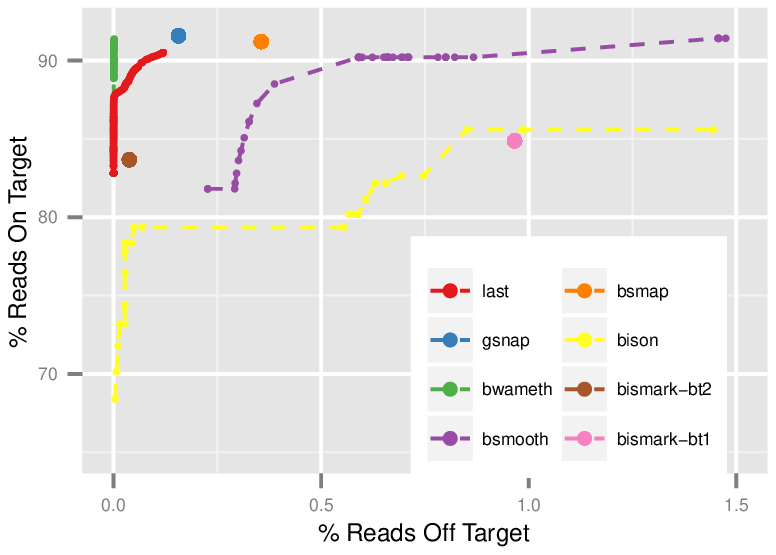}}
    \caption{Percent of paired-end, \emph{trimmed}, \emph{error-free, simulated} 100-base reads on (y) and off (x) target for the tested aligners. Aligners that report mapping quality are shown as connected dots for each quality cut-off. Reads are limited to those considered as primary, mapped alignments by the aligner.
}\label{suppfig:06}
\end{figure}

\subsection{Resources}

\begin{table}[H]
    \centering
    \caption{Resources on error-free simulated data
(programs run with more processes will use more memory)}
    \begin{tabular}{lllll} \hline
    trimmed & program & CPU time(min) & mem(GB) & dataset \\ \hline

no &    bis1 & 61.05 & 8.27 & error-free \\
no &    bis2 & 289.67 & 9.46 & error-free \\
no &    bison & 179.17 & 9.27 & error-free \\
no &    bsmap & 66.03 & 22.72 & error-free \\
no &    bsmooth & 492.62 & 9.37 & error-free \\
no &    bwa & 170.66 & 21.11 & error-free \\
no &    gsnap & 48.36 & 10.25 & error-free \\
no &    last & 214.68 & 25.90 & error-free \\

yes &    bis1 & 64.32 & 8.27 & error-free \\
yes &    bis2 & 245.42 & 9.48 & error-free \\
yes &    bison & 184.55 & 9.10 & error-free \\
yes &    bsmap & 67.68 & 22.94 & error-free \\
yes &    bsmooth & 481.99 & 9.38 & error-free \\
yes &    bwa & 173.93 & 21.43 & error-free \\
yes &    gsnap & 51.11 & 11.69 & error-free \\
yes &    last & 138.03 & 25.90 & error-free \\

    \end{tabular}
\end{table}

\section{Improved Accuracy for Stranded Capture Experiments}

Below we show a reduction in the percent of off-target reads by
considering only the strand targetted by our capture protocol.

\begin{figure}[H]
    \centerline{\includegraphics[width=86mm]{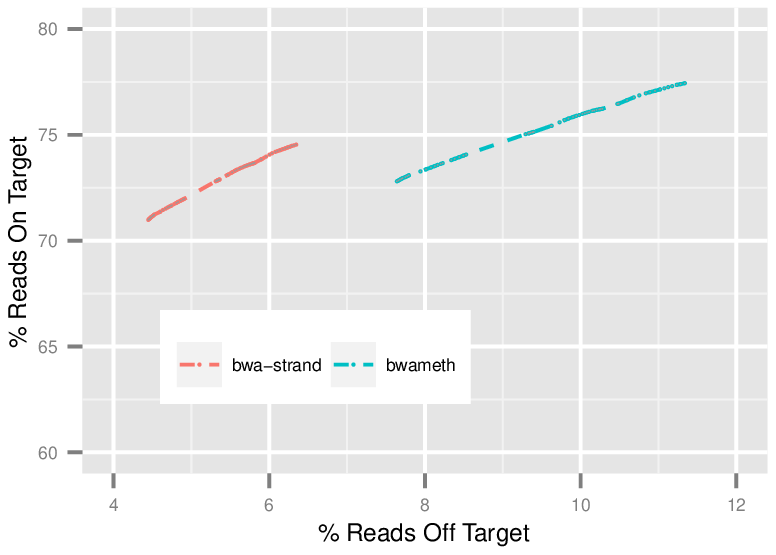}}
    \caption{For a capture experiment that targets only one strand, we can
    improve accuracy by considering only reads that map to that strand.
    Here, we show the effect from the original bwameth alignments as shown
    in Figure 1 compared to the alignments restricted to the strand targeted
    by the capture method.}\label{suppfig:07}
\end{figure}

This is implemented in \textit{bwa-meth} via the \emph{--set-as-failed}
flag so that reads mapping to a given strand are given the SAM flag
indicating that they failed vendor quality control checks. Invocation
would look like this:

\begin{lstlisting}[language=bash]
bwameth.py \
    --reference $REF \
    --set-as-failed f \
    $READS_R1 $READS_R2

\end{lstlisting}
to set reads mapping to the forward, or original top strand
as failed.

\section{Mapping and Trimming}
We note that trimming improves accuracy for most aligners but the
difference is very small for \textit{bwa-meth}.

\section{\textit{bwa-meth} Installation And Requirements}

\textit{Bwa-meth} depends on samtools and a single python library, \textit{toolshed}.
The latter can be installed by running \emph{sudo python setup.py install} from the main
directory of the \textit{bwa-meth} project. Samtools is a C library installed on most
systems and available at https://github.com/samtools/samtools.
Further installation instructions are available in the \textit{bwa-meth} README at
https://github.com/brentp/bwa-meth/

For tabulation of methylation by CpG, \emph{Bis-SNP} \cite{bissnp} is required.
The java .jar file is available from: http://sourceforge.net/projects/bissnp/files/

For CNV detection from BS-Seq data, the R package \emph{cn.mops} \cite{cnmops} is required.
It is available
from bioconductor \cite{bioconductor} at: http://bioconductor.org/packages/devel/bioc/html/cn.mops.html
\section{Additional Features}
\textit{Bwa-meth} and GSNAP are the only programs that output a BAM file that passes picard's ValidateSam without errors.
Last does not report the proper pair information in all cases and none of the other aligners add a read-group.
For the comparison, we added sorting and forced SAM output by the other aligners regardless of their default.
\textit{Bwa-meth} outputs a read-group for each sample by default and allows that to be customized.

\textit{bwa-meth} can calculate bias of methylation estimates by location in the read:

\begin{lstlisting}[language=bash]
python bias-plot.py input.bam ref.fa
\end{lstlisting}
This will create a bias-plot showing bases in the reads that should not be considered
for methylation calls.

\textit{Bwa-meth} defers tabulation of the methylation scores to Bis-SNP \cite{bissnp} by offering a simplified interface:
\begin{lstlisting}[language=bash]
bwameth.py tabulate \
    --trim 3,3 \
    --map-q 30 \
    --bissnp BisSNP-0.82.2.jar \
     --reference /path/to/ref.fasta \
     input.bam
\end{lstlisting}
Where the arguments are sent to Bis-SNP to, for example trim the first and last 3 bases
from each read to avoid bias.

A full example on real data is at:
https://github.com/brentp/bwa-meth/tree/master/example/

\bibliographystyle{abbrv}
    \bibliography{document}